# Numerical Modeling in Wave Scattering Problem for Small Particles

M. I. Andriychuk

Institute of Applied Problems of Mechanics
and Mathematics, NASU
Lviv, Ukraine
andr@iapmm.lviv.ua

A. G. Ramm

Mathematics Department
Kansas State University
Manhattan, KS, USA
ramm@math.ksu.edu

*Abstract*—A numerical solution to the problem of wave scattering by many small particles is studied under the assumption $ka \ll 1$, $d \gg a$, where $a$ is the size of the particles and $d$ is the distance between the neighboring particles. Impedance boundary conditions are assumed on the boundaries of small particles. The results of numerical simulation show good agreement with the theory. They open a way to numerical simulation of the method for creating materials with a desired refraction coefficient.

*Keywords-component; wave scattering; small particle; asymptotical approach, numerical modeling*

## I. INTRODUCTION

Theory of wave scattering by small particles of arbitrary shapes was developed by A. G. Ramm in [1] (see also [2]), where analytical formulas for $S$-matrix for acoustic and electromagnetic wave scattering by small bodies are derived. The proposed approach allows one to create the materials with a desired spatial dispersion, i.e., one can create the refraction coefficient $n^2(x,\omega)$ with a desired $\omega$-dependence, where $\omega$ is the frequency. In particular, one can create materials with negative refraction. Such materials are of interest in many applications, see, e.g., [3]-[7].

The proposed theory allows one to calculate the $S$-matrix with arbitrary accuracy. This can be used in many practical electromagnetic problems. An asymptotically exact solution of the many-body wave scattering problem was given in [8] under the assumptions $ka \ll 1$, $d = O(a^{1/3})$, $M = O(1/a)$, where $a$ is the size of the particles, $k = 2\pi/\lambda$ is the wave number, $d$ is the distance between neighboring particles, and $M$ is the total number of the particles embedded in a bounded domain $D \subset R^3$. An impedance boundary condition on the boundary $S_m$ of the $m$-th particle $D_m$ was assumed. In [9] the assumptions are more general:

$$\zeta_m = \frac{h(x_m)}{a^\kappa}, \quad d = O(a^{(2-\kappa)/3}), \quad M = O(\frac{1}{a^{2-\kappa}}), \quad \kappa \in (0,1), \quad (1)$$

where $\zeta_m$ is the boundary impedance, $h_m = h(x_m)$, $x_m \in D_m$, and $h(x) \in C(D)$ is an arbitrary continuous in $\overline{D}$ function, $\operatorname{Im} h \leq 0$.

The initial field $u_0$ satisfies the Helmholtz equation in $R^3$ and the scattered field satisfies the radiation condition. We assume in this paper that $\kappa \in (0,1)$ and the small particle $D_m$ is a ball of radius $a$ centered at the point $x_m$, $1 \leq m \leq M$.

## II. THE SOLUTION OF SCATTERING PROBLEM

The scattering problem is

$$[\nabla^2 + k^2 n_0^2(x)]u_M = 0 \text{ in } R^3 \setminus \bigcup_{m=1}^M D_m, \quad (2)$$

$$\frac{\partial u_M}{\partial N} = \zeta_m u_M \text{ on } S_m, \ 1 \leq m \leq M, \quad (3)$$

where

$$u_M = u_0 + v_M, \quad (4)$$

$u_0$ is solution to (2), (3) with $M = 0$, (i.e. in the absence of embedded particles) and with the incident field $e^{ik\alpha \cdot x}$, and $v_M$ satisfies the radiation condition.

It was proved in [8] that unique solution to (2) - (3) is of the form

$$u_M(x) = u_0(x) + \sum_{m=1}^M \int_{S_m} G(x,y)\sigma_m(y)dy, \quad (5)$$

where $G(x,y)$ is Green's function of Helmholtz equation in $R^3$, i.e., in the case when $M = 0$.

Let us define the "effective field" $u_e$, acting on the $m$-th particle:

$$u_e(x) := u_e(x,a) := u_e^{(m)}(x) := u_M(x) - \int_{S_m} G(x,y)\sigma_m(y)dy,$$

$$x \in R^3. \quad (6)$$

The function $\sigma_m(y)$ solves an exact integral equation, which is solved asymptotically in [9] as $a \to 0$. Let $h(x) \in C(D)$, $\text{Im } h \leq 0$, be arbitrary. Let $\Delta_p \subset D$ be any subdomain of $D$, and $N(\Delta_p)$ be the number of particles in $\Delta_p$. We assume that

$$N(\Delta_p) = \frac{1}{a^{2-\kappa}} \int_{\Delta_p} N(x) dx [1 + o(1)], \quad a \to 0, \quad (7)$$

where $N(x) \geq 0$ is a given continuous function in $D$. The following result was proved in [9] (Theorem 1): there exists the limit $u(x)$ of $u_e(x)$ as $a \to 0$:

$$\lim_{a \to 0} \| u_e(x) - u(x) \|_{C(D)} = 0, \quad (8)$$

and $u(x)$ solves the following equation:

$$u(x) = u_0(x) - 4\pi \int_D G(x, y) h(y) N(y) u(y) dy. \quad (9)$$

This is the derived in [9] equation for the limiting effective field in the medium, created by embedding many small particles with the distribution law (7).

### III. APPROXIMATE REPRESENTATION OF EFFECTIVE FIELD

Let us derive an explicit formula for the effective field $u_e$. Rewrite the exact formula (5) as:

$$u_M(x) = u_0(x) + \sum_{m=1}^{M} G(x, x_m) Q_m +$$

$$+ \sum_{m=1}^{M} \int_{S_m} [G(x, y) - G(x, x_m)] \sigma_m(y) dy, \quad (10)$$

where

$$Q_m = \int_{S_m} \sigma_m(y) dy. \quad (11)$$

Using some estimates of $G(x, y)$ [8] and the asymptotic formula for $Q_m$ from [9], we can rewrite the exact formula (10) as follows:

$$u_M(x) = u_0(x) + \sum_{m=1}^{M} G(x, x_m) Q_m + o(1),$$

$$a \to 0, \quad |x - x_m| \geq a. \quad (12)$$

The number $Q_m(x)$ is given by the asymptotic formula

$$Q_m = -4\pi h(x_m) u_e(x_m) a^{2-\kappa} [1 + o(1)], \quad a \to 0, \quad (13)$$

and the asymptotic formula for $\sigma_m$ is

$$\sigma_m = -\frac{h(x_m) u_e(x_m)}{a^\kappa} [1 + o(1)], \quad a \to 0. \quad (14)$$

Finally, formula for $u_e(x)$, $|x - x_j| \sim a$, is:

$$u_e^{(j)}(x) = u_0(x) - 4\pi \sum_{m=1, m \neq j}^{M} G(x, x_m) h(x_m) u_e(x_m) a^{2-\kappa} \times$$

$$\times [1 + o(1)]. \quad (15)$$

Equation (9) for the limiting effective field $u(x)$ is used in numerical calculations when the number $M$ is very large, say $M = 10^b, b \geq 5$.

In order to verify the applicability of formula (15) for practical applications, it is necessary to investigate the behavior of the solution to equation (9) and compare it with $u_e(x)$.

### IV. REDUCTION OF THE SCATTERING PROBLEM TO SOLVING LINEAR ALGEBRAIC SYSTEMS

The numerical calculation of the field $u_e$ by formula (15) requires the knowledge of the numbers $u_m := u_e(x_m)$. These numbers are obtained by solving a linear algebraic system (LAS). This system is:

$$u_j = u_{0j} - 4\pi \sum_{m=1, m \neq j}^{M} G(x_j, x_m) h(x_m) u_m a^{2-\kappa},$$

$$j = 1, 2, ..., M. \quad (16)$$

This LAS is convenient for numerical calculations, because its matrix is often diagonally dominant. It follows from the results in [10], that for sufficiently small $a$ this system is uniquely solvable.

For finding the solution to the limiting equation (9), we use the collocation method from [10], which yields the following LAS:

$$u_j = u_{0j} - 4\pi \sum_{p=1, m \neq j}^{P} G(x_j, x_p) h(y_p) N(y_p) u(y_p) |\Delta_p|,$$

$$p = 1, 2, ..., P, \quad (17)$$

where $P$ is number of small cubes $\Delta_p$, $y_p$ is center of $\Delta_p$, $|\Delta_p|$ is volume of $\Delta_p$. We assume that the union of $\Delta_p$ forms a partition of $D$, and the diameter of $\Delta_p$ is $O(d^{1/2})$.

From the computational point of view solving LAS (17) is much easier than LAS (16), because $P \ll M$.

We have two different LAS, corresponding to formula (15) and to equation (9). Solving these LAS, one can compare their solutions and evaluate the limits of applicability of the asymptotic approach from [9] to solving many-body wave scattering problem in the case of small particles.

## V. NUMERICAL CALCULATIONS

The following numerical experiments are important from the practical point of view:
a) for not very large $M$, say, $M$ =2, 5, 10, 25, 50, one wants to investigate what the values of $a$ and $d$ are for which the asymptotic formula (12) (without the remainder $o(1)$) started to be not applicable;
b) for large $M$, say, $M = 10^5$, $M = 10^6$, one wants to investigate the relative difference between the solution to the limiting equation (9) and the solution to (16);
c) investigation of the relative difference between the solution to the limiting equation (9) and the solution to (17);
d) investigation of the relative difference between the solution to (16) and (17);
e) for given $n^2(x)$ and $n_0^2(x)$ when one uses Ramm's method for creating materials, one wants to investigate what the smallest $M$ (or, equivalently, largest $a$) is for which the corresponding $n^2_{M(x)}$ differs from the desired $n^2(x)$ by not more than, say, 5% - 10%.

We deal here with the numerical calculations related to item a), i.e., with checking the applicability of the asymptotic theory in the case of small $M$. The calculations are carried out for $k = 1$, $\kappa = 0.9$, and $N(x) = N = \text{const}$. The accuracy of the solution to LAS (16) and LAS (17) corresponding to the limiting equation (9) is tested for various values of $a$ and fixed values of $N$. The solution of limiting equation (9) and corresponding LAS (17) is obtained by the collocation method [10]. It was proven in [10] that the solution to LAS (17) tends to solution of limiting equation (9) when the number of collocation points tends to infinity. The relative error of its solution does not exceed 0.1% in the range of the parameters considered here; the number $P$ of collocation points sufficient for obtaining such accuracy, equals $11^3$.

In Fig. 1, the relative error of real (real), imaginary (imag) parts, and moduli (abs) of solutions to (16) and (17) is shown for the case of $M = 4$ particles; the distance between them is determined as $d = a^{(2-\kappa)}C$, where $C$ is additional parameter of optimization (in our case $C = 5$), $N(x) = 5$. The minimal relative error of the both solutions does not exceed 0.05% and is reached for the range $a$ from 0.02 to 0.03. The value of function $N(x)$ influences the quality of approximation. The relative error for $N(x) = 40$ with other parameters unchanged is shown in Fig. 2. The error is smallest at $a = 0.01$, and it grows when $a$ increases. The minimal accessible error for this case is equal to 0.01%.

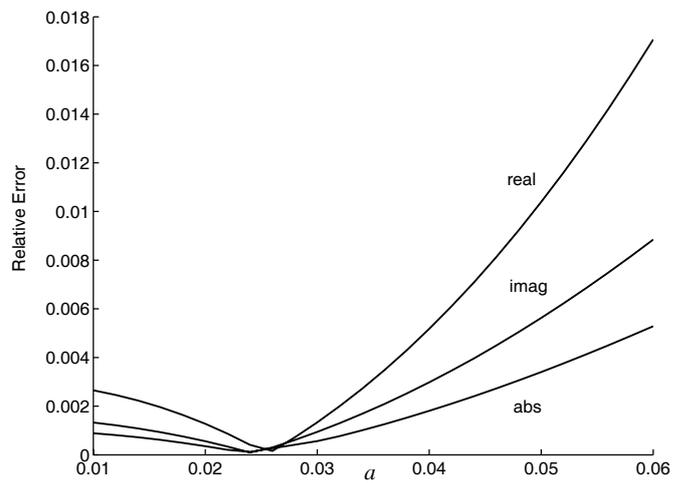

Fig.1. Relative error of solution versus the size $a$ of particle, $N(x) = 5$.

The error dependence on the distance $d$ between the particles at the fixed $a$ was investigated too. In Fig. 3, the relative error versus the parameter $d$ is shown. The number of particles $M = 4$, the radius of particle $a = 0.01$. The minimal error is reached at $C = 14$. The error grows significantly at smaller and greater values of $d$. Similar results are shown for $a = 0.02$ in Fig. 4. The error is more sensitive to change of distance $d$ for this case. The minimal value of error for last case is equal to 0.05%, and it is reached at $C = 8$.

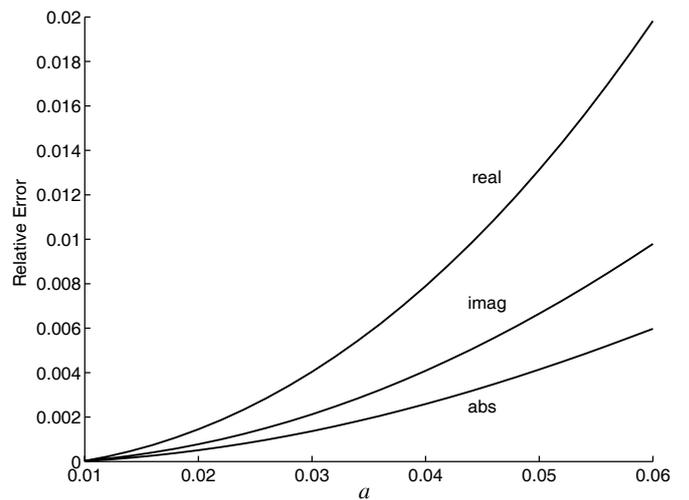

Fig.2. Relative error of solution versus the size $a$ of particle, $N(x) = 40$.

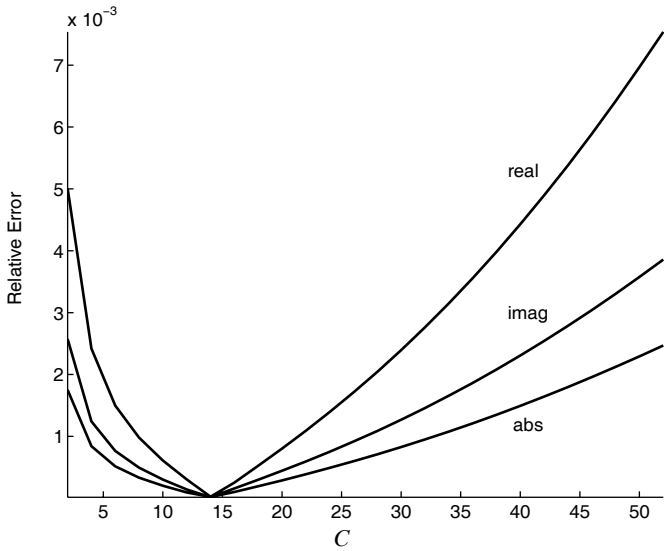

Fig. 3. Relative error of solution versus the distance $d$ between particles, $a = 0.01$, $d = a^{2-\kappa}C$.

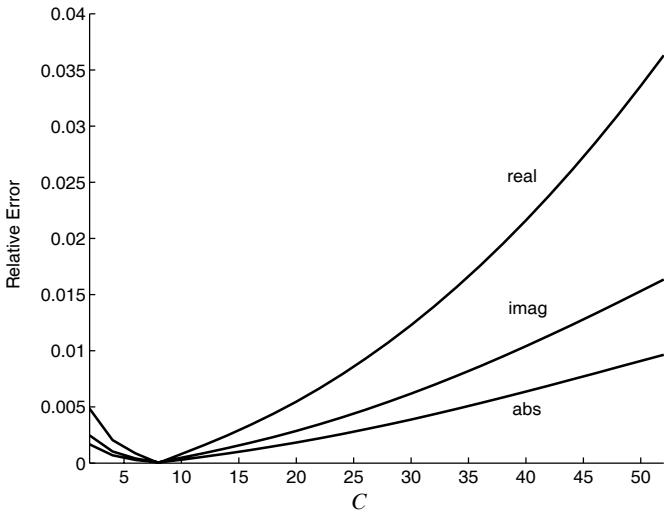

Fig. 4. Relative error of solution versus the distance $d$ between particles, $a = 0.02$, $d = a^{2-\kappa}C$.

The calculations show that the accuracy of approximation depends essentially on $a$. The minimal error is 1% at $a = 0.04$.

The optimal values of $d$ are given for small $M$ in Table I, and for $M \in (10, 40)$ in Table II respectively. The approximate values of $d$ are shown here. The numerical results demonstrate that in order to decrease the relative error of solution to system (16), it is necessary to make $a$ smaller.

TABLE I. OPTIMAL VALUES OF $d$ FOR SMALL $M$

|  | $M$ value | | | |
|---|---|---|---|---|
|  | $M = 2$ | $M = 4$ | $M = 6$ | $M = 8$ |
| $a = 0.01$ | 0.038 | 0.025 | 0.026 | 0.027 |
| $a = 0.02$ | 0.053 | 0.023 | 0.027 | 0.054 |

TABLE II. OPTIMAL VALUES OF $d$ FOR MEDIUM $M$

|  | $M$ value | | | |
|---|---|---|---|---|
|  | $M = 10$ | $M = 20$ | $M = 30$ | $M = 40$ |
| $a = 0.01$ | 0.011 | 0.010 | 0.007 | 0.0063 |
| $a = 0.02$ | 0.016 | 0.018 | 0.020 | 0.023 |

## VI. CONCLUSIONS

The numerical results based on the asymptotical approach to solving the scattering problem in media with many small particles demonstrate a possibility to apply the proposed technique for modelling the bodies with desired properties, in particular, with a prescribed refraction coefficient. The respective optimization problem can be formulated and solved using the results received in the process of solving the considered scattering problem. The elaborated software tools allow to solve the both problems in a wide range of geometrical and physical parameters of problem.